\documentclass[12pt,a4paper]{conference}

\usepackage{fancyhdr}
\usepackage{graphicx,amsmath,amssymb,cite}
\usepackage{multind}
\makeindex{author} \makeindex{subject}

\newcommand{\beq}{\begin{equation}}
\newcommand{\eeq}[1]{\label{#1}\end{equation}}
\newcommand{\eeqn}{\end{equation}}

\newcommand{\beqa}{\begin{eqnarray}}
\newcommand{\eeqa}[1]{\label{#1}\end{eqnarray}}
\newcommand{\eeqan}{\end{eqnarray}}

\let\bar=\overbar

\newcommand{\Dslash}{\not{\hbox{\kern-4pt $D$}}}
\newcommand{\dslash}{\not{\hbox{\kern-2pt $\del$}}}

\newcommand{\msb}{{\bar{\ssstyle M \kern -1pt S}}}

\begin{document}
%%%%%%%%%%%%%%%%%%%%%%%%%%%%%%%%%%%%%%%%%%%%%%%%%%%%%%%%%%%%%%%%%%%%%%%

\Chapter{Studies of the $\Lambda(1405)$ in Proton-Proton
Collisions with ANKE at COSY-J\"ulich}
           {Studies of $\Lambda(1405)$ with ANKE@COSY}
{I.~Zychor}
\vspace{-6 cm}\includegraphics[width=6 cm]{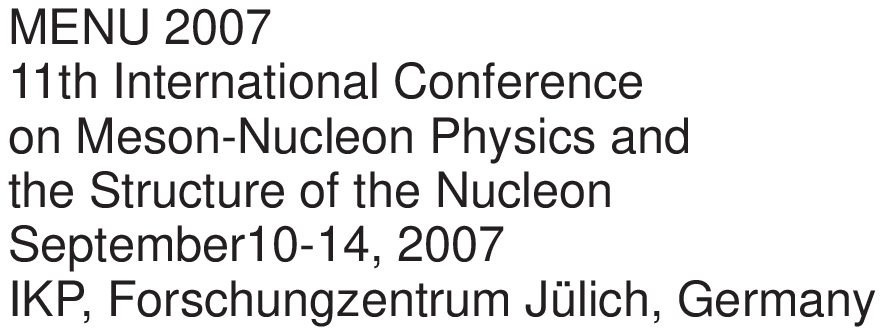}
%\bigskip\bigskip
\vspace{4 cm}

\addcontentsline{toc}{chapter}{{\it I.~Zychor}} \label{authorStart}
%%%%%%%%%%%%%%%%%%%%%%%%%%%% NEW SWITCHES %%%%%%%%%%%%%%%%%%%%%%%%%%%%%%

\begin{raggedright}

{\it I.~Zychor
\index{author}{Zychor, I.}}\\
The Andrzej So{\l}tan Institute for Nuclear Studies\\
05-400 \'Swierk, Poland\\
\bigskip\bigskip
\end{raggedright}

\begin{center}
\textbf{Abstract}
\end{center}

The lineshape of the $\Lambda(1405)$ was studied in the
$pp\rightarrow pK^+ Y^0$ reaction at a beam momentum of
3.65\,GeV/c at COSY-J\"ulich. The ANKE spectrometer was used to
identify two protons, one positively charged kaon, and one
negatively charged pion in the final state. Invariant--mass and
missing--mass techniques were applied to separate two neighbouring
neutral excited hyperon resonances, the $\Sigma^0(1385)$ and
$\Lambda(1405)$. Both the shape and the position of the
$\Lambda(1405)$ distribution are similar to those measured in
other reactions and this information contributes to the ongoing
debate regarding the structure of this resonance.

\section{Introduction}

The $\Lambda(1405)$ is a well established four--star
resonance~\cite{PDG} but it is still not well understood as a
baryonic state; it does not fit in easily within the simple quark
picture~\cite{Isgur}. The $\Lambda(1405)$ might be the
spin-multiplet partner of the $J^P=\frac{3}{2}^-$ $\Lambda(1520)$,
a meson--baryon resonance, a $\bar K N$ quasibound
state~\cite{DalitzTuan}, or a $q^4 \bar q$ pentaquark
state~\cite{Inoue}. Recent theoretical investigations based on
chiral dynamics predict the existence of two poles in the vicinity
of the $\Lambda(1405)$~\cite{Jido, Magas, Geng} with a decay
spectrum that depends upon the production process. In any event,
the $\Lambda(1405)$ does not have a Breit--Wigner shape because of
the opening at 1432~MeV/c$^2$ of the decay mode
$\bar{K}N$~\cite{Hem, Dalitz,Thomas}. Independent of the model, if
the $\Lambda(1405)$ were a single quantum state, its lineshape
should be independent of the method of production.

The $\Sigma^0(1385)$ and $\Lambda(1405)$ resonances overlap
significantly because their widths of 36\,MeV/c$^2$ and
50\,MeV/c$^2$, respectively, are much larger than the mass
difference of $\sim$\,20\,MeV/c$^2$. This is the main experimental
difficulty in investigating the $\Lambda(1405)$ nature \emph{via}
the $\Sigma^+\pi^-$ and $\Sigma^-\pi^+$ decay modes since these
are also possible final states for the $\Sigma^0(1385)$
disintegration. However, the $\Lambda(1405)\to\Sigma^0 \pi^0$
decay can be used to identify this resonance unambiguously because
isospin forbids this mode for the $\Sigma^0(1385)$.

\begin{figure}[ht]
\begin{center}
\includegraphics[width=5.5 cm]{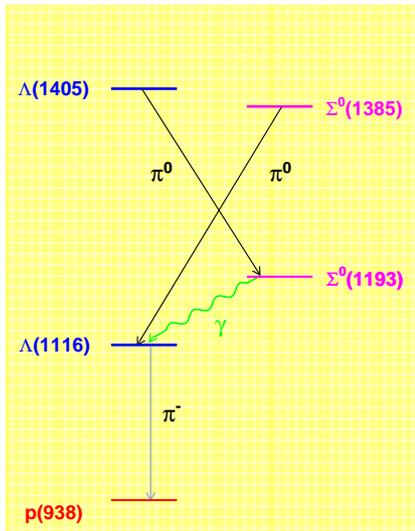}
\caption{Simplified decay scheme for the $\Lambda(1405)$ and
$\Sigma^0(1385)$ hyperon resonances} \label{fig:decay_modes}
\end{center}
\end{figure}

In Fig.~\ref{fig:decay_modes} the simplified decay scheme of
excited neutral resonances with masses below 1432~MeV/c$^2$
demonstrates the differences between $\Sigma^0(1385)$ and
$\Lambda(1405)$ utilised in the present analysis.

%%%%%%%%%%%%%%%%%%%%%%%%%%%%%%%%%%%%%%%%%%%%%%%%%%%%%%%%%%%%%%%%%%%%%%%%%%%%5
%%%%%%%%%%%%%%%%%%%%%%%%%%%%%%%%%%%%%%%%%%%%%%%%%%%%%%%%%%%%%%%%%%%%%%%%%%%%%

\section{Experiment, analysis and simulations}

The experiment was performed at the Cooler Synchrotron COSY, a
medium energy accelerator and storage ring for protons and
deuterons, which is operated at the Research Center J\"ulich
(Germany)~\cite{COSY}. COSY supplied a stored proton beam with a
momentum of 3.65~GeV/c at a revolution frequency of
\mbox{$\sim10^{6}\,\textrm{s}^{-1}$}. Using a hydrogen
cluster--jet target, the average luminosity during the
measurements was $L=(55\pm 8)\,\textrm{pb}^{-1}$.

The ANKE spectrometer~\cite{ANKE_NIM} used in the experiments
consists of three dipole magnets that guide the circulating COSY
beam through a chicane. The central C--shaped spectrometer dipole
D2, placed downstream of the target, separates the reaction
products from the beam. The ANKE detection system, comprising
range telescopes, scintillation counters and multi--wire
proportional chambers, registers simultaneously positively and
negatively charged particles and measures their
momenta~\cite{K_NIM}.

The following configuration of detectors was used to measure
particles over a particular momentum range:
\begin{enumerate}
\item forward (Fd) and side--wall (Sd) counters for protons
between 0.75\,GeV/c and the kinematic limit, %
\item telescopes and side--wall scintillators for $K^+$ between
0.2 and 0.9\,GeV/c, %
\item scintillators for $\pi^-$ between 0.2 and 1.0\,GeV/c.
\end{enumerate}
The angular acceptance of the spectrometer dipole D2 is
$|\vartheta_{\mathrm H}|\lesssim 12^{\circ}$ horizontally and
$|\vartheta_{\mathrm V}| \lesssim 5^{\circ}$ vertically. Momenta,
reconstructed from tracks in multi--wire proportional chambers,
allow the masses of particles to be determined to within
$\sim10\,$MeV/c$^2$.

A multiparticle final state, containing two protons, a positively
charged kaon, a negatively charged pion and an unidentified
residue $X^0$ selected the \mbox{$pp \rightarrow pK^+ p \pi^-
X^0$} reaction. In the $\Sigma^0(1385) \rightarrow \Lambda \pi^0$
decay the $X^0$ residue is a $\pi^0$ while, for the $\Lambda(1405)
\rightarrow \Sigma^0 \pi^0$ decay, $X^0 = \pi^0 \gamma$ (see
Fig.~\ref{fig:decay_modes}).

%%%%%%%%%%%%%%%%%%%%%%%%%%%%%%%%%%%%%%%%%%%%%%%%%%%%%%%%%%%%%%%%%%%%%%%%%%%%5
\begin{figure}[hb]
\begin{center}
\includegraphics[width=6.5 cm]{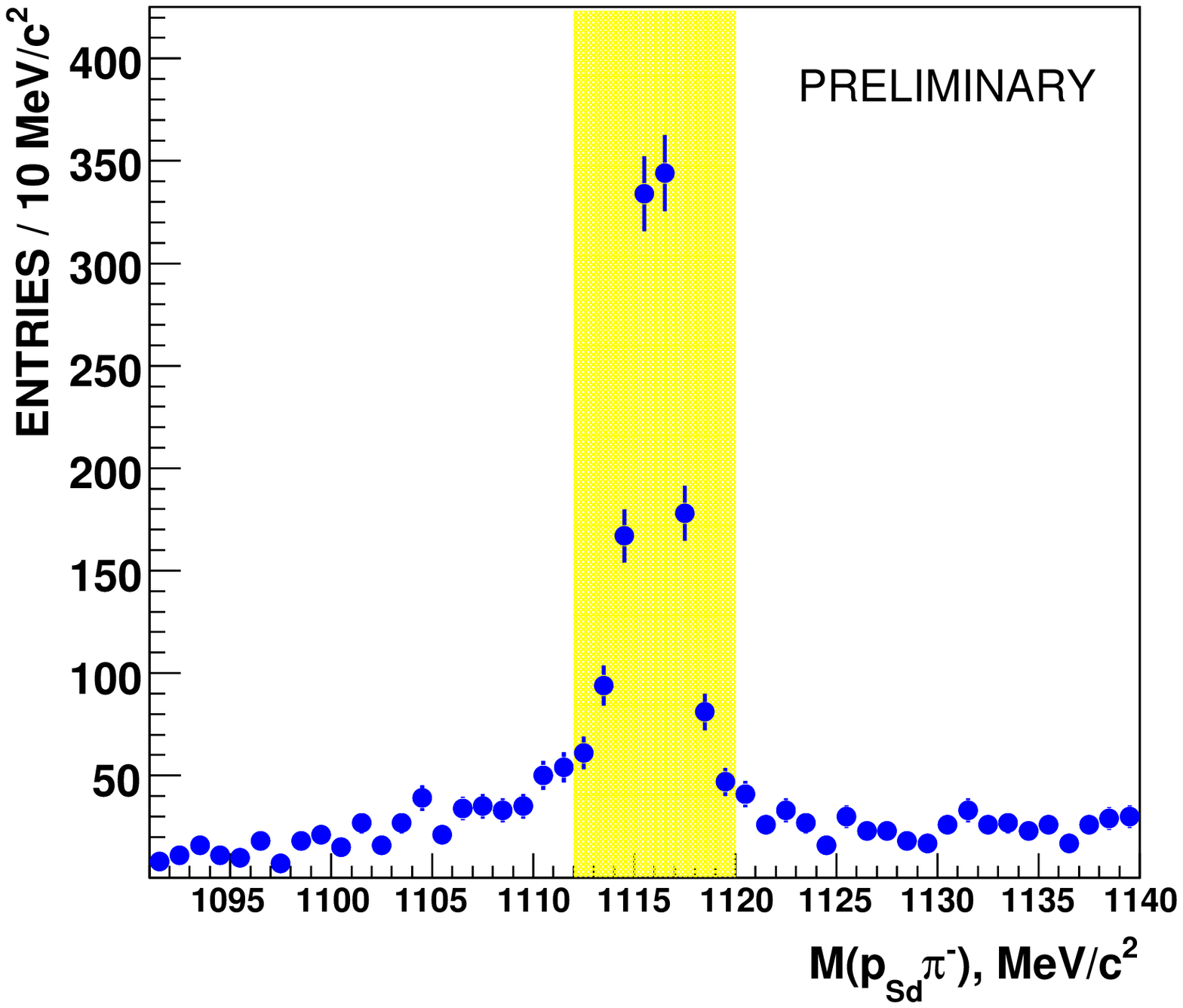}
\hspace{0.1 cm}
\includegraphics[width=6.5 cm]{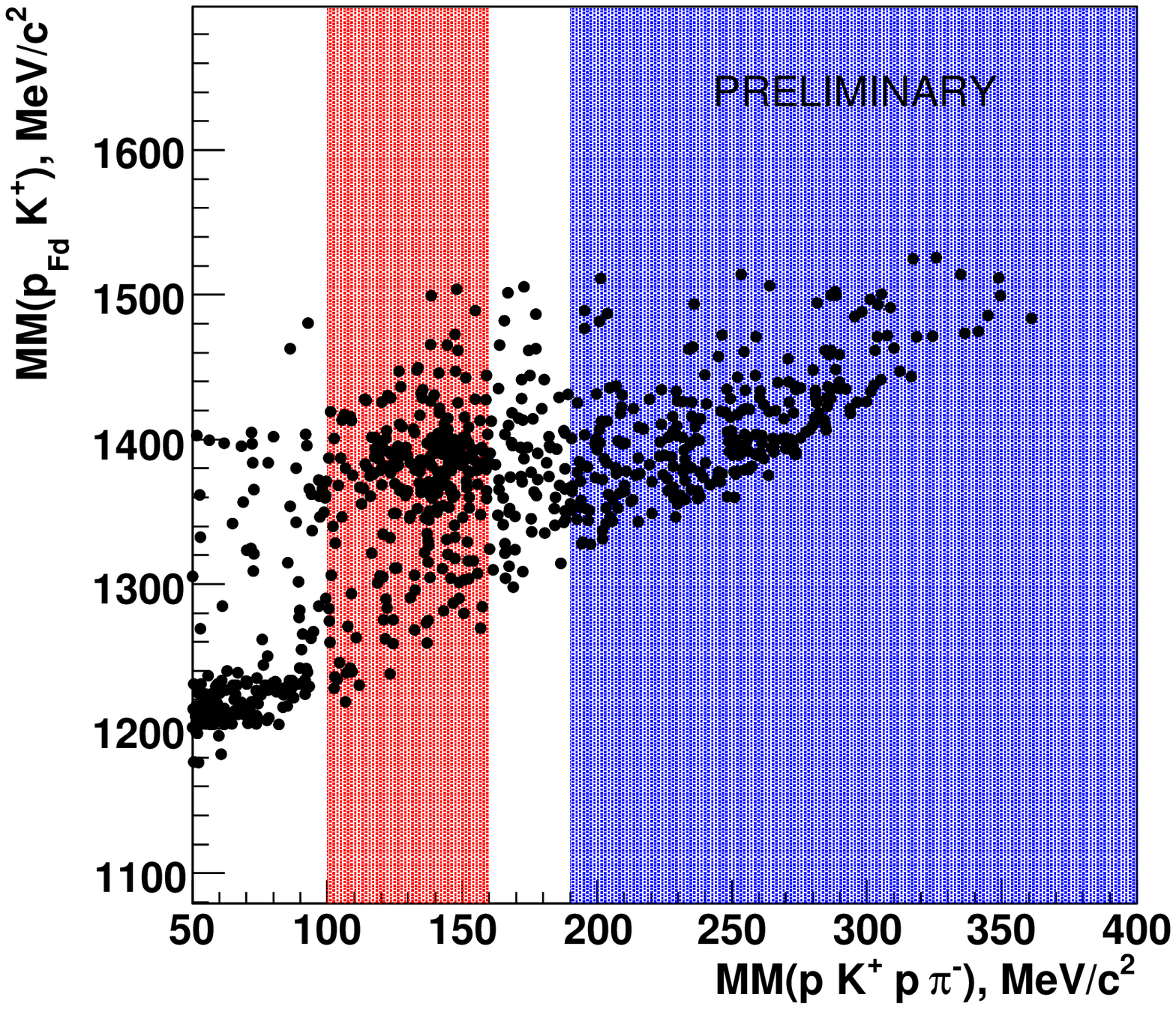}
\caption{ \textit{\underline{Left:}} Invariant mass
$M(p_{Sd}\pi^-)$ measured in the 3.65\,GeV/c $pp \rightarrow pK^+
Y^0$ reaction. The yellow horizontal box shows the band used to
select the $\Lambda$. \textit{\underline{Right:}} Missing mass
$MM(p_{Fd}K^+)$ \textit{versus} the missing mass $MM(pK^+\pi^-
p)$. The left (red) vertical box covers the $\pi^0$ region and the
right (blue) one has $MM(p K^+\pi^- p) > 190\,\textrm{MeV}/c^2\gg
m(\pi^0)$.} \label{fig:inv_MM}
\end{center}
\end{figure}
%%%%%%%%%%%%%%%%%%%%%%%%%%%%%%%%%%%%%%%%%%%%%%%%%%%%%%%%%%%%%%%%%%%%%%%%%%%%%

The following method was used to separate the $\Lambda(1405)$ from
the $\Sigma^0(1385)$:
\begin{enumerate}
\item identify four particles: $p_{Fd}$, $p_{Sd}$, $K^+$ and
$\pi^-$, %
\item analyse events with the invariant mass of the
$p_{Sd} \pi^-$ pair equal to the mass of the~$\Lambda$, %
\item select events with the missing mass of ($p_{Fd}$, $p_{Sd}$,
$K^+$, $\pi^-$) equal to a $\pi^0$ mass to isolate the
$\Sigma^0(1385)$ and much higher than the $\pi^0$ mass to identify
the $\Lambda(1405)$.
\end{enumerate}

In the left part of Fig.~\ref{fig:inv_MM} the invariant mass
$M(p_{Sd}\pi^-)$ of the $p_{Sd} \pi^-$ pairs is shown, where the
protons were registered in the side-detector counters. In the mass
region around 1116\,MeV/c$^2$ a peak with a FWHM of $\sim
5\,$MeV/c$^2$ is visible on a background that is mostly
combinatorial in nature. The vertical box marks invariant--masses
between 1112 and 1120\,MeV/c$^2$. Events within this box are
plotted in the right panel of Fig.~\ref{fig:inv_MM} in a
distribution of $MM(p_{Fd}K^+)$ \textit{versus} $MM(p K^+\pi^-
p)$. The two vertical bands show the four--particle missing--mass
$MM(p K^+\pi^- p)$ criteria used to separate the $\Sigma^0(1385)$
candidates from those of the $\Lambda(1405)$. The left band is
optimised to identify a $\pi^0$ whereas the right one selects
masses significantly greater than $m(\pi^0)$. The deviation of
$\sim 8\,$MeV/c$^2$ from the nominal pion mass of 135\,MeV/c$^2$
is not unexpected in a mass reconstruction involving four
particles.

%%%%%%%%%%%%%%%%%%%%%%%%%%%%%%%%%%%%%%%%%%%%%%%%%%%%%%%%%%%%%%%%%%%%%%%%%%%%5
\begin{figure}[ht]
\begin{center}
\includegraphics[width=6.5 cm]{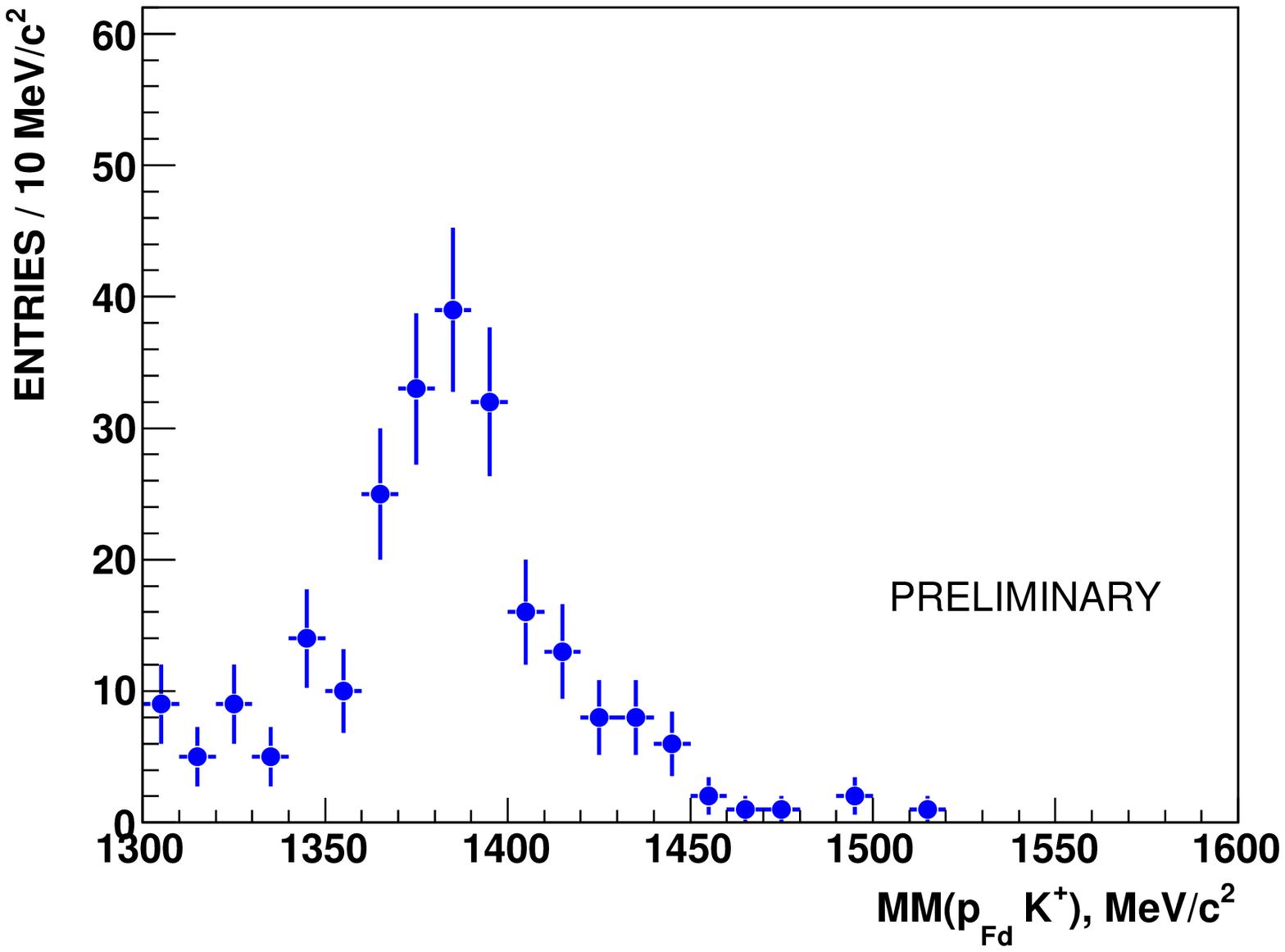}
\hspace{0.1 cm}
\includegraphics[width=6.5 cm]{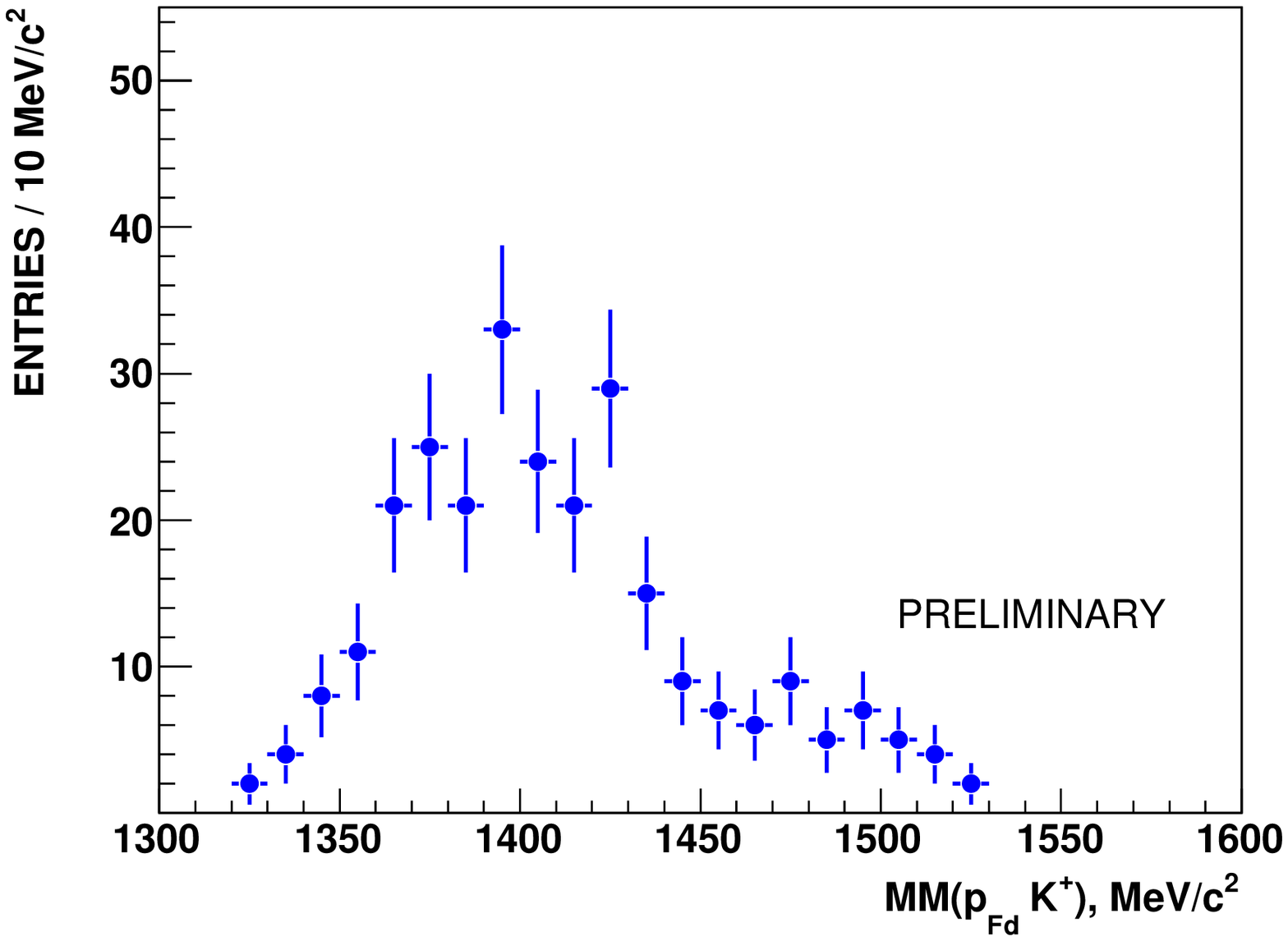}
\caption{Missing--mass $MM(p_{Fd}K^+)$ distribution for the
$pp\rightarrow pK^+ p \pi^- X^0$ reaction for events with
$M(p_{Sd}\pi^-)\approx m(\Lambda)$. The distribution obtained for
$MM(p K^+\pi^- p)\approx m(\pi^0)$ is presented in the left panel
and for $MM(p K^+\pi^-p)> 190\,\textrm{MeV/c}^2$ in the right one.
} \label{fig:mm2_1385_1405}
\end{center}
\end{figure}
%%%%%%%%%%%%%%%%%%%%%%%%%%%%%%%%%%%%%%%%%%%%%%%%%%%%%%%%%%%%%%%%%%%%%%%%%%%%%

In the left part of Fig.~\ref{fig:mm2_1385_1405} the missing--mass
$MM(p_{Fd} K^+)$ distribution is shown for $MM(p K^+\pi^- p)
\approx m(\pi^0)$. A peak around a mass of 1385~MeV/c$^2$ and a
width of $\sim50\,$MeV/c$^2$ is seen on a rather small background.
In the right part of Fig.~\ref{fig:mm2_1385_1405} the
distribution, obtained for $MM(p K^+\pi^- p) >190\,$MeV/c$^2$, has
a peak near 1400\,MeV/c$^2$ and a tail on the high missing--mass
side.

%%%%%%%%%%%%%%%%%%%%%%%%%%%%%%%%%%%%%%%%%%%%%%%%%%%%%%%%%%%%%%%%%%%%%%%%%%%%5
\begin{figure}[ht]
\begin{center}
\includegraphics[width=7.5 cm]{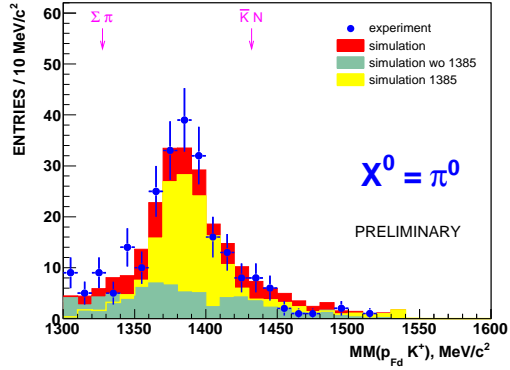}
\caption{ Missing--mass $MM(p_{Fd}K^+)$ distribution for the
$pp\rightarrow pK^+ p \pi^- X^0$ reaction for events with
$M(p_{Sd}\pi^-)\approx m(\Lambda)$ and $MM(p K^+\pi^- p)\approx
m(\pi^0)$. Experimental points with statistical errors are
compared to the red histogram of the fitted overall Monte Carlo
simulations. The simulation includes resonant contributions
(yellow) and non--resonant phase--space production (green). Arrows
indicate the $\Sigma \pi$ and $\bar{K} N$ thresholds.}
\label{fig:mm2_exp_MC}
\end{center}
\end{figure}
%%%%%%%%%%%%%%%%%%%%%%%%%%%%%%%%%%%%%%%%%%%%%%%%%%%%%%%%%%%%%%%%%%%%%%%%%%%%%

In order to explain the measured spectra, Monte Carlo simulations
were performed to estimate backgrounds from non--resonant and
resonant reactions. The following non--resonant processes have
been included:
\begin{enumerate}
\item $pp \rightarrow NK^+\pi X (\gamma)$
\item $pp \rightarrow NK^+\pi\pi X (\gamma)$
\end{enumerate}
with $X$ representing any allowed $\Lambda$ or $\Sigma$ hyperon.
The second group consists of the following exclusive hyperon
production reactions:
\begin{enumerate}
\item $pp \rightarrow pK^+ \Sigma^0(1385)$
\item $pp \rightarrow pK^+ \Lambda(1405)$
\item $pp \rightarrow pK^+ \Lambda(1520)$
\end{enumerate}
The simulations, based on the GEANT3 package, were performed in a
similar manner to those in Ref.~\cite{Y1480}.

In the study of $\Sigma^0(1385)$ production and its backgrounds,
events were generated according to phase space using a
relativistic Breit--Wigner parameterisations for the known hyperon
resonance~\cite{PDG}. The relative contributions of the resonant
and non--resonant reactions were deduced by fitting the
experimental data to the simulated spectra. In
Fig.~\ref{fig:mm2_exp_MC} the histograms show the resonant
contribution from the $pp \rightarrow pK^+ \Sigma^0(1385)$
reaction (solid-yellow) and the sum of non--resonant contributions
(solid-green). The result of the overall simulations is shown as a
red histogram.

\begin{figure}[ht]
\begin{center}
\includegraphics[width=5.7 cm]{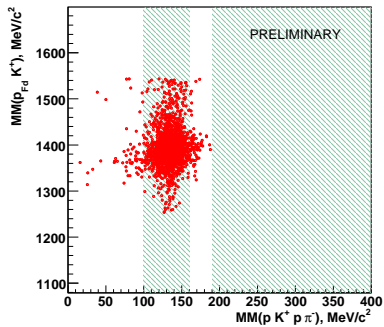}
\caption{ Simulated distribution of events with missing mass
$MM(p_{Fd}K^+)$ \textit{versus} $MM(pK^+\pi^- p)$. The left shaded
vertical box covers the $\pi^0$ region and the right one has
$MM(pK^+\pi^-p)> 190\,\textrm{MeV/c}^2\gg m(\pi^0)$. Notice an
absence of events in the right box. } \label{fig:MC_1385_for_1405}
\end{center}
\end{figure}

Turning now to the $\Lambda(1405)$, simulations show that the
$\Sigma^0(1385)$ does not contaminate the missing--mass $MM(p
K^+\pi^- p)$ range above 190\,MeV/c$^2$ (see
Fig.~\ref{fig:MC_1385_for_1405}). This point is crucial since it
allows us to obtain a clean separation of the $\Sigma^0(1385)$
from the $\Lambda(1405)$.

%%%%%%%%%%%%%%%%%%%%%%%%%%%%%%%%%%%%%%%%%%%%%%%%%%%%%%%%%%%%%%%%%%%%%%%%%%%%5
%%%%%%%%%%%%%%%%%%%%%%%%%%%%%%%%%%%%%%%%%%%%%%%%%%%%%%%%%%%%%%%%%%%%%%%%%%%%%

\begin{figure}[ht]
\begin{center}
\includegraphics[width=5.7 cm]{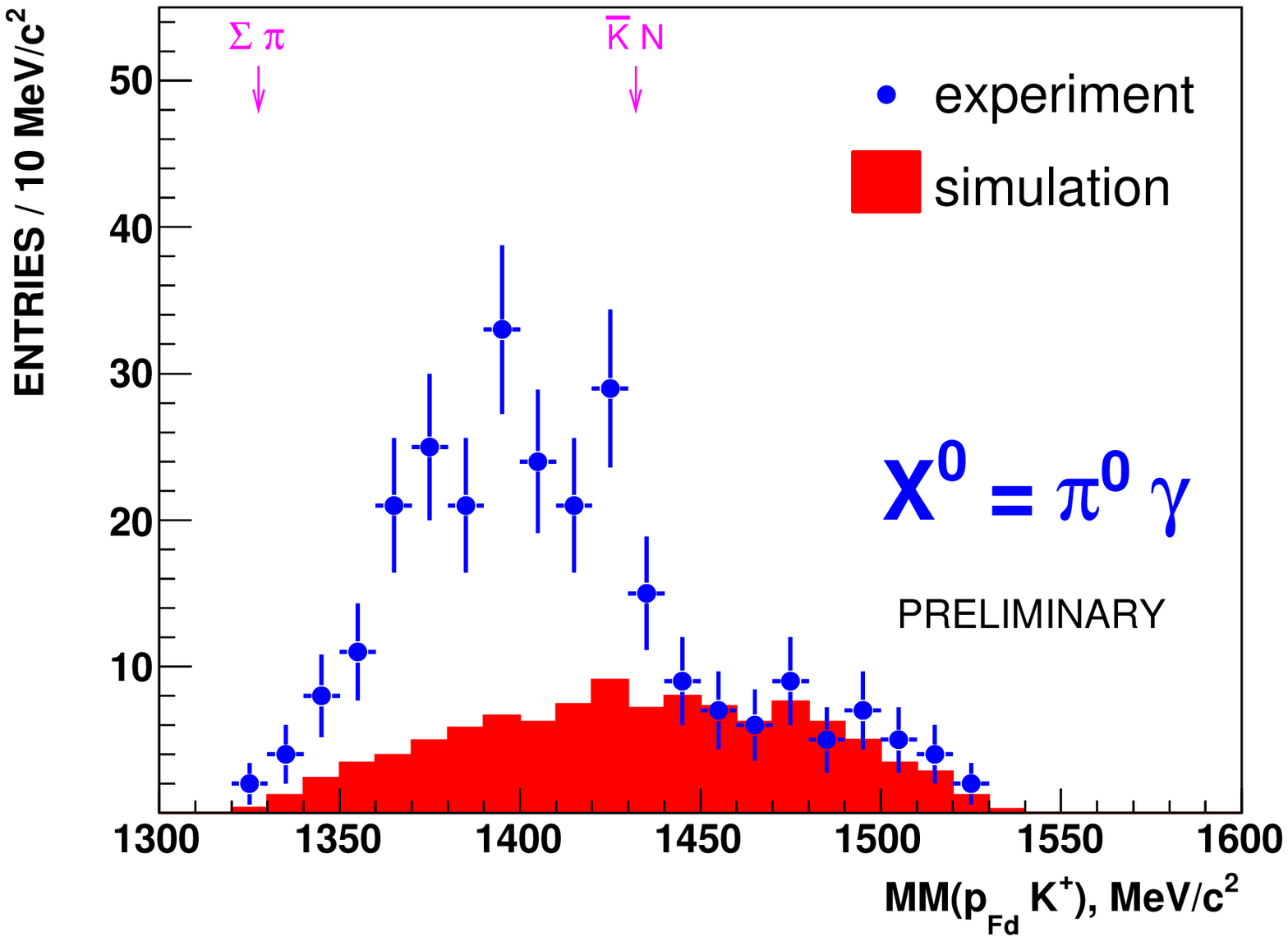} \hspace*{0.1cm}
\includegraphics[width=5.7 cm]{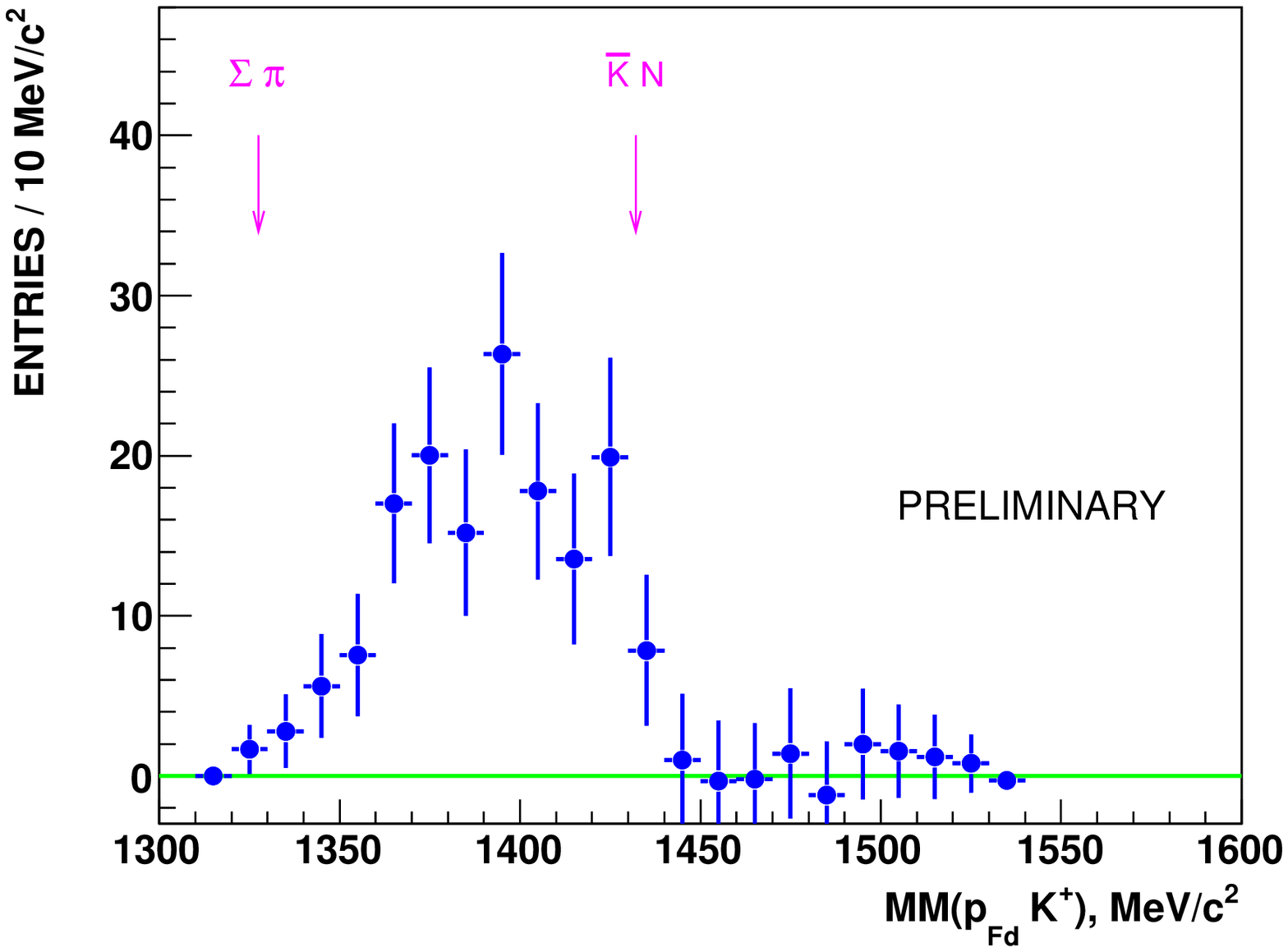}
\caption{
\textit{\underline{Left:}} Experimental missing--mass $MM(p_{Fd}K^+)$ distribution for
the $pp \rightarrow pK^+p \pi^- X^0$ reaction for
events with $M(p_{Sd}\pi^-) \approx m(\Lambda)$ and $MM(p
K^+\pi^- p) > 190\,\textrm{MeV/c}^2$
compared to the shaded histogram of the fitted
non--resonant Monte Carlo simulation.
\textit{\underline{Right:}} The
background--subtracted lineshape of the $\Lambda(1405)$ decaying
into $\Sigma^0 \pi^0$.
}
\label{fig:fig_1405}
\end{center}
\end{figure}

In order to extract the $\Lambda(1405)$ distribution from the
measured $\Sigma^0 \pi^0$ decay, the non--resonant contributions
have first been fitted to the experimental data. The resulting
non--resonant background is indicated by the shaded histogram in
the left panel of Fig.~\ref{fig:fig_1405}. When this is subtracted
from the data, we obtain the distribution shown as experimental
points in the right panel of Fig.~\ref{fig:fig_1405}.

%%%%%%%%%%%%%%%%%%%%%%%%%%%%%%%%%%%%%%%%%%%%%%%%%%%%%%%%%%%%%%%%%%%%%%%%%%%%5
%%%%%%%%%%%%%%%%%%%%%%%%%%%%%%%%%%%%%%%%%%%%%%%%%%%%%%%%%%%%%%%%%%%%%%%%%%%%%

\section{Results}

In Table~\ref{ta1} the information that is relevant for the
evaluation of the total cross section is given. For both the
hyperons measured this is of the order of a few $\mu \textrm{b}$.

\begin{table}[phb]
\caption{Total cross section for the production of the $\Sigma^0(1385)$ and $\Lambda(1405)$
resonances in the 3.65\,GeV/c $pp \rightarrow  pK^+ Y^0$ reaction}
\begin{tabular}{l|l|l} \hline \hline
                          & $\Sigma^0(1385)$    & $\Lambda(1405)$  \\ \hline
number of events          & $170\pm 26$        & $156 \pm 23$  \\
acceptance                & $2.0\times10^{-6}$ & $4.4 \times 10^{-6}$   \\
combined BR (\%)          & 56                  & 21  \\
luminosity (pb$^{-1}$)    & $55\pm 8$          & $55 \pm 8$   \\
detection efficiency (\%) & $55 \pm 11$        & $55 \pm 11$  \\
\hline cross section ($\mu\textrm{b}$)& $5.0 \pm
1.2_{\text{stat}}\pm 2.0_{\text{syst}}$ &$ 5.6 \pm
0.8_{\text{stat}}\pm 2.2_{\text{syst}}$  \\ \hline \hline
\end{tabular} \label{ta1}
\end{table}

The $(\Sigma \pi)^0$ invariant--mass distributions have been
previously studied in two hydrogen bubble chamber experiments.
Thomas \textit{et al.}~\cite{Thomas} found $\sim$\,400
$\Sigma^+\pi^-$ or $\Sigma^-\pi^+$ events corresponding to the
$\pi^- p \rightarrow K^0 \Lambda(1405)\rightarrow K^0 (\Sigma
\pi)^0$ reaction at a beam momentum of 1.69\,GeV/c.
Hemingway~\cite{Hem} used a 4.2\,GeV/c kaon beam to investigate
$K^- p \rightarrow \Sigma^+(1660) \pi^- \rightarrow \Lambda(1405)
\pi^+ \pi^- \rightarrow  (\Sigma ^+ \pi^-) \pi^+ \pi^-$ and
measured 1106 events~\cite{Hem}.

\begin{figure}[ht]
\begin{center}
\includegraphics[width=7.5 cm]{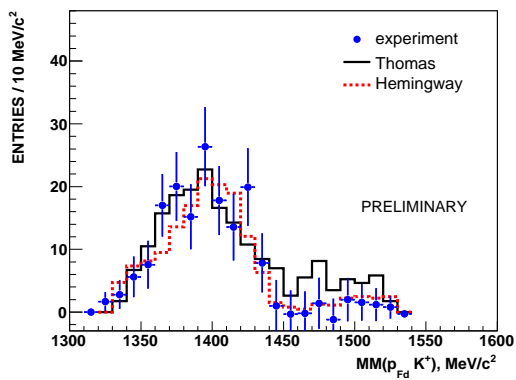}
\caption{ The background--subtracted lineshape of the $\Lambda(1405)$ decaying
into $\Sigma^0 \pi^0$ (points) compared to $\pi^- p
\rightarrow K^0 (\Sigma \pi)^0$~\cite{Thomas} (black--solid
line) and $K^-p\rightarrow \pi^+\pi^-\Sigma^+\pi^-$~\cite{Hem}
(red--dotted line) data.}
\label{fig:fig_1405_comp}
\end{center}
\end{figure}

In Fig.~\ref{fig:fig_1405_comp} our experimental points
are compared to the results of Thomas and Hemingway, which have
been normalised by scaling their values down by factors of $\sim$3
and $\sim$7, respectively. The effect of the $\bar{K} N$ threshold
is quite obvious in these data, with the $\Lambda(1405)$ mass distribution being
strongly distorted by the opening of this channel.
Despite the very different production mechanisms, the three
distributions have consistent shapes.

%%%%%%%%%%%%%%%%%%%%%%%%%%%%%%%%%%%%%%%%%%%%%%%%%%%%%%%%%%%%%%%%%%%%%%%%%%%%5
%%%%%%%%%%%%%%%%%%%%%%%%%%%%%%%%%%%%%%%%%%%%%%%%%%%%%%%%%%%%%%%%%%%%%%%%%%%%%

This might suggest that, if there are two states present in this
region, then the reaction mechanisms in the three cases are
preferentially populating the same one. It should, however, be
noted that by identifying a particular reaction mechanism, the
proponents of the two--state solution can describe the shape of
the distribution that we have found~\cite{Geng}.

\section{Outlook}

The decay of excited hyperons $Y^{0*}$ \textit{via} $\Lambda
\pi^0$ and $\Sigma^0 \pi^0 \rightarrow \Lambda \gamma \pi^0$ can
be detected directly in electromagnetic calorimeters by
registering neutral particles, \textit{i.e.}\ $\gamma$ and/or
$\pi^0$. Measurements of such channels are underway in $\gamma
p$~reactions (CB/TAPS~at~ELSA~\cite{ELSA},
SPring$-$8/LEPS~\cite{LEPS}) and are also planned in $pp$
collisions with WASA~at~COSY~\cite{WASA}.

\index{subject}{hyperon}
\index{subject}{resonance}
\index{subject}{hadronic reaction}
\index{subject}{$\Lambda(1405)$}
\index{subject}{ANKE}
\index{subject}{COSY}

\section*{Acknowledgments}

The results presented constitute the common effort of many members
of the ANKE collaboration (www.fz-juelich.de/ikp/anke) and the COSY accelerator group, as
described in Ref.~\cite{L1405}. This work has been supported by
COSY-FFE Grant, BMBF, DFG and Russian Academy of Sciences.

\end{document}